\documentclass[]{svjour3}                     

\usepackage{mathrsfs}

\usepackage[dvips,final]{graphicx}
\usepackage[dvips]{geometry} 
\usepackage{color} 
\usepackage{epsfig} 
\usepackage{latexsym}
\usepackage{pstricks}

\usepackage[hyphens]{url}
\usepackage{algorithm}
\usepackage{algorithmic}
\usepackage{amssymb, amsmath}
\usepackage{mathrsfs}
\usepackage{natbib}

\newcommand{\NP}{{\cal NP}}

\def \NP {{$\cal NP$}}

\begin{document}

\title{
A note on the paper ``Minimizing total tardiness on parallel machines with preemptions''
by~\cite{KW10}
} 

\author{D. Prot \and O. Bellenguez-Morineau \and C. Lahlou}
\institute{D. Prot \and O. Bellenguez-Morineau \and C. Lahlou\at
Ecole des Mines de Nantes, rue Alfred Kastler, 44307 Nantes Cedex\\
\email{damien.prot, odile.morineau, chams.lahlou@mines-nantes.fr}}

\date{Received: date / Accepted: date}

\maketitle

\begin{abstract} 
In this note, we point out two major errors in the paper 
``Minimizing total tardiness on parallel machines with preemptions''
by~\cite{KW10}. More precisely, they proved that both problems
$P|pmtn|\sum T_j$ and $P|r_j,p_j=p,pmtn|\sum T_j$ are \NP-Hard.
We give a counter-example to their proofs, letting the complexity
of these two problems open.


\end{abstract}

\section{Reductions used in~\cite{KW10}}
In~\cite{KW10}, the authors propose a reduction of $P|pmtn|\sum T_j$ 
(in Section~3) and $P|r_j,p_j=p,pmtn|\sum T_j$ (in Section~4)
from {\sc Partition}: given a set of positive integers $a_1,\dots,a_k,b$ with
$\sum_{i=1}^k a_i = 2b$, does there exist a subset $I\subset \{1,\dots,k\}$
such that $\sum_{i\in I}a_i = b$?

Given an instance of {\sc Partition}, we only detail the reduction
of  $P|pmtn|\sum T_j$ since it is used in the proof of the \NP-Hardness of $P|r_j,p_j=p,pmtn|\sum T_j$.
The instance of $P|pmtn|\sum T_j$ is composed of $2k^2+k+1$ jobs and $k$ machines,
and a constant $L=\frac{4kb^3+2b}{k}$ is used.
The authors define three classes of jobs:
\begin{itemize}
\item
the $a-$jobs: it is composed of $k$ jobs $a_i$ with $p_i=a_i$ and $d_i=L$, $i\in\{1,\dots,k\}$.
\item
the $ba-$jobs: it is composed of $2k^2$ jobs, $2k$ equivalent jobs $ba_i$ with $p_i=b^2a_i$ and $d_i=L-a_i$,
$i\in\{1,\dots,k\}$.
\item
one long job $b^3$, with processing time $b^3$ and due date $b^3$.
\end{itemize}
The authors claim that {\sc Partition} has a solution if and only if there exists a schedule
with $\sum T_j \leq b^3+b$.
The necessary part of this result is quite obvious.
The sufficient part is more complex, and the authors claim that 
if there exists a schedule with  $\sum T_j \leq b^3+b$, then the
set of $ba-$jobs completed after time point $L$ defines the solution for {\sc Partition}.
We show in the next section that this sufficient part
does not hold, i.e. starting with a solution of $P|pmtn|\sum T_j$ such
that  $\sum T_j \leq b^3+b$ may not lead to a solution of {\sc Partition}.

\section{A counter-example to the reduction and its consequences}
The counter-example to the reduction is an instance of {\sc Partition} with $k=3$:
$a_1=1$, $a_2=2$, and $a_3=3$. The corresponding instance of $P|pmtn|\sum T_j$ is
hence composed of $22$ jobs and $3$ machines, and the constant $L$ is equal to $110$.
The jobs have the following characteristics:
\begin{itemize}
\item
the $a-$jobs: $a_1$,$a_2$ and $a_3$ have a processing time of $1$, $2$, and $3$ and a common due date $L=110$.
\item
the $ba-$jobs: there are six jobs $ba_1$ of processing $9$ and due date $109$,
six jobs $ba_2$ of processing $18$ and due date $108$,
six jobs $ba_3$ of processing $27$ and due date $107$.
\item
one long job $b^3$ with processing $27$ and due date $27$.
\end{itemize}

\begin{center}
\begin{figure}
\begin{center}
\includegraphics[angle=-90,scale=0.32]{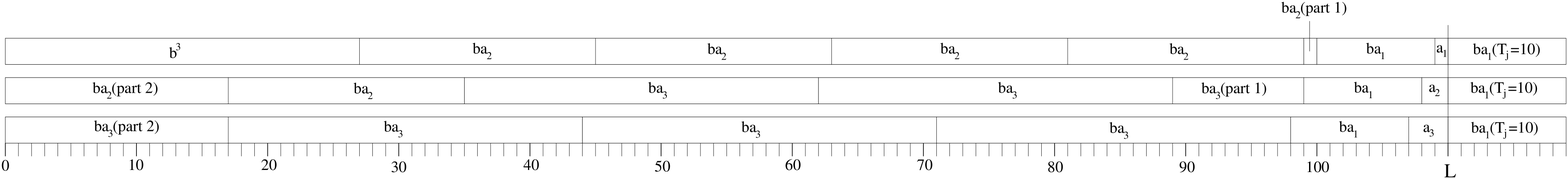}
\caption{A counter-example to the reduction}
\label{fig:KW10}
\end{center}
\end{figure}
\end{center}

A schedule such that $\sum T_j \leq b^3+b=30$ is proposed in Figure~\ref{fig:KW10}.
There are only three late jobs, of type $ba_1$, each of them finishing $10$ time units
after its due date. According to~\cite{KW10}, the corresponding solution of {\sc Partition}
is $I=\{a_1\}$, which is obviously wrong.

As a consequence, the proposed reduction of $P|pmtn|sum T_j$ from {\sc Partition}
does not hold; the reduction of $P|r_j,p_j=p,pmtn|\sum T_j$ from  {\sc Partition}
being based on the same construction, it is also wrong.
Hence, it is still an open question to know whether $P|pmtn|sum T_j$ and $P|r_j,p_j=p,pmtn|\sum T_j$
are \NP-Hard problems or not. 
\bibliographystyle{plainnat}
\bibliography{Biblio}

\begin{thebibliography}{1}
\providecommand{\natexlab}[1]{#1}
\providecommand{\url}[1]{\texttt{#1}}
\expandafter\ifx\csname urlstyle\endcsname\relax
  \providecommand{\doi}[1]{doi: #1}\else
  \providecommand{\doi}{doi: \begingroup \urlstyle{rm}\Url}\fi

\bibitem[Kravchenko and Werner(2010)]{KW10}
S.A. Kravchenko and F.~Werner.
\newblock Minimizing total tardiness on parallel machines with preemptions.
\newblock \emph{Journal of Scheduling}, pages 1--8, 2010.
\newblock URL \url{http://dx.doi.org/10.1007/s10951-010-0198-5}.
\newblock 10.1007/s10951-010-0198-5.

\end{thebibliography}
\end{document}